\documentclass[prd,aps,floats,twocolumn]{revtex4}

\usepackage[dvips]{graphicx}
\usepackage{amssymb}
\usepackage{amsmath}
\usepackage{ctable} 
\usepackage{caption}

\begin{document}

\title{Self-Similar One-Dimensional Quasilattices}

\author{Latham Boyle$^1$ and Paul J. Steinhardt$^{1,2,3}$}

\affiliation{$^1$Perimeter Institute for Theoretical Physics, \\ 
Waterloo, Ontario N2L 2Y5, Canada \\
$^2$Princeton Center for Theoretical Science, Princeton University \\
Princeton, NJ, 08544 USA \\
$^3$Department of Physics, Princeton University\\Princeton, NJ, 08544 USA}  

\begin{abstract}
We study 1D quasilattices, especially self-similar ones that can be used to generate two-, three- and higher-dimensional quasicrystalline
tesselations that have matching rules and invertible self-similar substitution rules (also known as inflation rules) analogous to the rules for generating Penrose tilings.  The lattice positions can be expressed in a closed-form  expression we call  {\it floor form}: $x_{n}=S(n-\alpha)+(L-S)\lfloor \kappa(n-\beta)\rfloor$, where $L >S>0$ and $0<\kappa<1$ is an irrational number.  We describe 
two equivalent geometric constructions of these quasilattices and show how they can be subdivided into various types of equivalence classes:  (i) {\it lattice equivalent}, where any two quasilattices in the same lattice equivalence class may be derived from one another by a local decoration/gluing rule; (ii) {\it self-similar}, a proper subset of lattice equivalent where, in addition, the two quasilattices are locally isomorphic; and (iii) {\it self-same}, a proper subset of self-similar where, in addition, the two quasilattices are globally isomorphic ({\it i.e.} identical up to rescaling).  For all three types of equivalence class, we obtain the explicit transformation law between the floor form expression for two quasilattices in the same class.  We tabulate (in Table I and Figure 5) the ten special self-similar 1D quasilattices relevant for constructing Ammann patterns and Penrose-like tilings in two dimensions and higher, and we explicitly construct and catalog the corresponding self-same quasilattices.  
\end{abstract}

\maketitle


\section{Introduction}

\begin{figure}
  \begin{center}
    \includegraphics[width=3.0in]{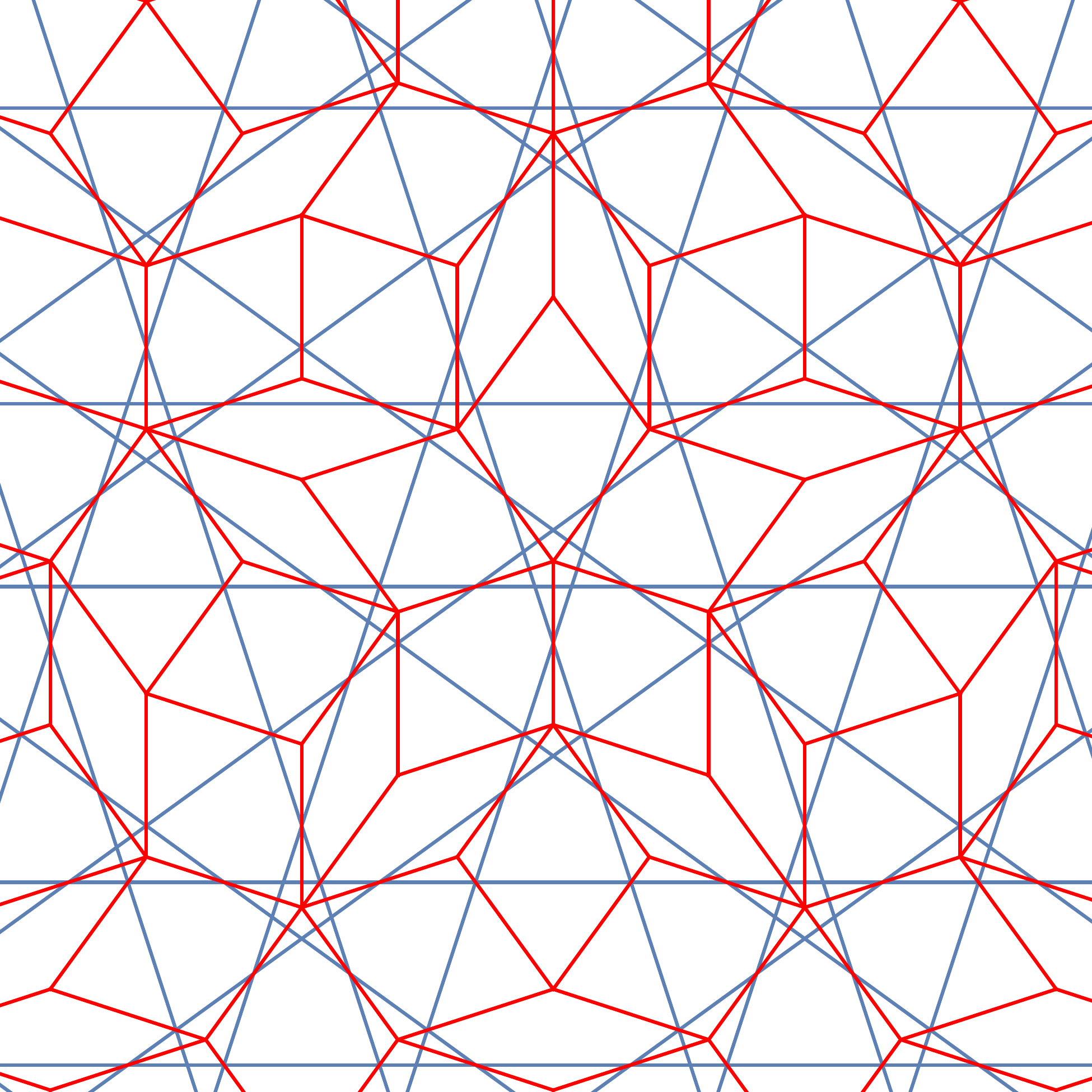}
  \end{center}
  \caption{The red lines show a portion of a Penrose tiling (constructed from two tiles -- a thin rhomb and a fat rhomb), while the blue lines show the corresponding Ammann pattern.}
 \label{AmmannPenrose}
\end{figure}

Penrose tilings \cite{Penrose74, Gardner77, Penrose78} were the inspiration for introducing the concept of quasicrystals \cite{LevineSteinhardt84} and have stimulated enormous progress in our understanding of aperiodic order in mathematics and physics \cite{GrunbaumShephard, Janot, Senechal, Baake2002, SteurerDeloudi, BaakeGrimme}.   These tilings exhibit a fascinating set of interrelated properties, including: (i) quasiperiodic translational order; (ii) crystallographically-forbidden 10-fold orientational order; (iii) discrete scale invariance (as embodied in so-called ``inflation/deflation" rules \cite{Gardner77}); (iv)  ``matching" rules that constrain the way two tiles can join edge-to-edge such that the tiles can only fill the plane by forming perfect Penrose tilings; and (v) a distinctive class of topological (``decapod") defects.   The Penrose tiles also have another important feature: the two tiles can each be decorated with a certain pattern of line segments that join together in a perfect Penrose tiling to form five infinite sets of parallel lines oriented along the five edges of a pentagon.  The lines  are spaced according to a 1D quasiperiodic sequence of long and short intervals called a ``Fibonacci quasilattice" (see Fig.~\ref{AmmannPenrose}).    The five sets of 1D quasilattices together form an Ammann pattern, named after Robert Ammann, who found this decoration \cite{GrunbaumShephard, SenechalOnAmmann}.  

In this paper we lay the 1D foundation for a new approach to Penrose tilings (and other objects like them, but with different symmetries and in higher dimensions) \cite{BoyleSteinhardtHigherD}.  The perspective developed and applied in \cite{BoyleSteinhardtHigherD} is that a Penrose-like tiling should be regarded as the dual of a more fundamental object: an Ammann pattern; and this Ammann pattern, in turn, can be derived from the relationship between two naturally-paired irreducible reflection groups (which we call a ``Coxeter pair").

Our focus in this paper is the analysis of the 1D quasilattices that serve as the building blocks for the Ammann patterns in higher dimensions.   Although our ultimate purpose is higher-dimensional quasicrystal tilings as described in \cite{BoyleSteinhardtHigherD}, the 1D quasilattices studied here are important objects in their own right (see {\it e.g.}\ \cite{Senechal, BaakeGrimme, deBruijn81a, BombieriTaylor1, BombieriTaylor2, Hof, Dyson1, Fogg, AlloucheShallit}), and a number of the new results about them that we present here are of independent interest.  Let us sketch the outline of this paper and highlight a few key results:

We begin, in Section II, by constructing the simplest class of 1D quasilattices: we will call them ``1D quasilattices of degree two" or ``quadratic 1D quasilattices".  
These are 1D quasiperiodic lattices constructed from just two intervals or ``tiles'' (call them $L$ and $S$, for ``long" and ``short"), with just two different separations between successive $L$'s, and just two different separations between successive $S$'s (the simplest possibility compatible with quasiperiodicity).  The quasilattice point positions can be specified by a closed-form analytic expression which has the basic ``floor form" $x_{n}=S(n-\alpha)+(L-S)\lfloor \kappa(n-\beta)\rfloor$ where $\lfloor x \rfloor$ denotes the ``floor of $x$" ({\it i.e.}\ the largest integer $\leq x$) and $L>S>0$, $\alpha$, $\beta$ and $0<\kappa<1$ are constants.   

The quasilattices can be constructed geometrically by first picking some input data: an arbitrary 2D lattice $\Lambda$, an integral basis $\{\vec{m}_{1},\vec{m}_{2}\}$ for $\Lambda$, and an arbitrary line $\vec{\;\!q}(t)$ that slices through the lattice with irrational slope.  We describe two equivalent constructions that produce the natural 1D quasilattice corresponding to this input data: (i) the first construction involves dualizing the ``1D bi-grid" obtained by intersecting the $\{\vec{m}_{1},\vec{m}_{2}\}$ integer grid lines with the line $\vec{\;\!q}(t)$; and (ii) the second construction is based on a cut-and-project scheme using $\Lambda$ as the lattice, $\vec{\;\!q}(t)$ as the ``cut" surface, and a parallelogram with sides $\vec{m}_{1}$ and $\vec{m}_{2}$ as the acceptance window.  At first glance, it might seem like the first (dualization) construction only defines the 1D quasilattice up to an overall (unfixed) translational phase ambiguity; but it will be important for later applications \cite{BoyleSteinhardtHigherD} to remove this phase ambiguity by imposing the condition that the 1D bi-grid is reflection symmetric if and only if the corresponding 1D quasilattice is reflection symmetric.  We then observe that this condition has a natural geometric interpretation in the second (cut-and-project) construction: it amounts to the requirement that, whenever the line $\vec{\;\!q}(t)$ intersects one of the $\{\vec{m}_{1},\vec{m}_{2}\}$ parallelograms in the lattice $\Lambda$, it is the {\it midpoint} of that parallogram that should be projected onto $\vec{\;\!q}(t)$ to define a point in the 1D quasilattice.  We show [see (\ref{xn_symmetric}) or (\ref{xn_asymmetric})] that these two geometric constructions yield 1D quasilattices captured by the floor-form expression described above; and, conversely, that any 1D quasilattice in floor-form (for any values of the parameters $S$, $L$, $\alpha$, $\beta$ and $\kappa$) can be obtained via these geometric constructions.

Section III concerns the following simple observation.  In Section II, we began by choosing a line $\vec{\;\!q}(t)$, a lattice $\Lambda$, and an integral basis $\{\vec{m}_{1},\vec{m}_{2}\}$ for $\Lambda$; and we obtained a corresponding 1D quasilattice $x_{n}$.  If we had instead chosen a different integral basis $\{\vec{m}_{1}',\vec{m}_{2}'\}$, we would have obtained a different quasilattice $x_{n}'$.  We will describe two such quasilattices as ``lattice equivalent."  The quasilattices $x_{n}$ and $x_{n}'$ might look quite dissimilar from one another in terms of their tile sizes and orderings, but (as we explain in Section III) each may be obtained from the other by a local ``substitution/gluing" rule that uses  the integer matrix $\tau$ to relate the old basis $\{\vec{m}_{1},\vec{m}_{2}\}$ to the new basis $\{\vec{m}_{1}',\vec{m}_{2}'\}$.  In this way, the set of quadratic 1D quasilattices is partitioned into ``lattice equivalence classes'' with a simple geometric interpretation.  Namely, the members of a given class correspond to the same line $\vec{\;\!q}(t)$ and the same lattice $\Lambda$, but different choices for the basis $\{\vec{m}_{1},\vec{m}_{2}\}$.

In Section IV, we identify the subset of quadratic 1D quasilattices that are {\it self-similar}.  For these lattices, there is a change of basis $\{\vec{m}_{1},\vec{m}_{2}\}\rightarrow\{\vec{m}_{1}',\vec{m}_{2}'\}$ that maps the quasilattice $x_{n}$ into a new quasilattice $x_{n}'$ that is not only lattice equivalent, but also {\it locally isomorphic} up to rescaling of the intervals between points.  For each self-similar 1D quasilattice, our construction identifies a canonical self-similar substitution/decoration rule, specifying not just the {\it number} of ``new" tiles which decorate each of the ``old" tiles, but also the particular order and phase of the new tiles in decorating the old.  This canonical substitution rule is always reflection symmetric.  We also obtain a simple and useful analytic expression for how the parameters in the floor-form expression for the ``old" quasilattice are related to the parameters in the floor-form expression for the ``new" quasilattice obtained from it by this canonical substitution rule.  In a generic (non-singular) self-similar quasilattice, the line $\vec{\;\!q}(t)$ does not intersect any of the points in the lattice $\Lambda$; but we also carefully treat the special (singular) case where $\vec{\;\!q}(t)$ does intersect a point in $\Lambda$, because the corresponding special quasilattices play an important role in the analysis of topological defects in Penrose-like tilings in two dimensions and higher \cite{BoyleSteinhardtDefects}.  Finally, since we are dealing with quadratic 1D quasilattices, the corresponding self-similar quasilattices are characterized by quadratic irrationalities.  In fact, only a small subset of these self-similar quadratic 1D quasilattices play a role as the building blocks for the Ammann patterns in two dimensions and higher \cite{BoyleSteinhardtHigherD}: the parameters and canonical substitution rules for these ten special quasilattices are presented in Table I and Figure 5.

In Section V, we identify the subset of quadratic 1D quasilattices that are not only self-similar under some $2\times2$ transformation $\tau$, but are exactly {\it $s$-fold self-same}; that is, $\tau^s$ maps the quasilattice $x_{n}$ to a new quasilattice $x_{n}'$ that is not merely locally-isomorphic, but actually {\it identical} to the original quasilattice (up to an overall rescaling).  We obtain a simple explicit formula for these $s$-fold self-same quasilattices, and also for the {\it number} of distinct $s$-fold self-same quasilattices.  These $s$-fold self-same quasilattices are naturally grouped into irreducible $s$-cycles: for each of the special quasi-lattices listed in Table I, we count the number of irreducible $s$-cycles, and list the results in Table II.  In comparing the results to the Online Encyclopedia of Integer Sequences (OEIS), some interesting connections appear.  These $s$-fold self-same 1D quasilattices, and irreducible $s$-cycles thereof, are the building blocks for $s$-fold self-same Ammann patterns and Penrose-like tilings in two dimensions and higher; and these, in turn, underlie a new scheme for discretizing scale invariant systems.

\section{Quadratic 1D quasilattices: two geometric perspectives}
\label{two_perspectives}

We will say that a 1D quasilattice is ``of degree two" or ``quadratic" 
if it can be described by the following ``floor form" expression:
\begin{equation}
  \label{second_order}
  x_{n}=S(n-\alpha)+(L-S)\lfloor\kappa(n-\beta)\rfloor.
\end{equation}
Here $\{L,S,\kappa,\alpha,\beta\}$ are real-valued constants (with $L>S>0$ and $0<\kappa<1$ irrational), $n$ is an integer that runs from $-\infty$ to $+\infty$, and $\lfloor x\rfloor$ is the ``floor" of $x$ {\it i.e.}\ the greatest integer $\leq x$.  Thus, as $n$ increases (from $N$ to $N+1$), $x_{n}$ correspondingly increases (from $x_{N}$ to either $x_{N+1}=x_{N}+L$ or $x_{N+1}=x_{N}+S$); in other words, Eq.~(\ref{second_order}) describes a sequence of isolated points along the real line, with just two different intervals between neighboring points: $L$ and $S$ (``long" and ``short").  The $L$'s and $S$'s form an infinite non-repeating sequence: the relative frequency with which $L$ and $S$ occur in the sequence is determined by $\kappa$ and the particular order in which they occur is determined by $\beta$, while $\alpha$ is an overall translation phase that determines where exactly the sequance is situated along the real line.

Note that quadratic 1D quasilattices are ``as simple as possible" in the sense that they are built from just two different intervals ($L$ and $S$); and, in addition, there are just two different separations between consecutive $S$'s, and just two different separations between consecutive $L$'s.  Anything simpler than this would be incompatible with quasiperiodicity.

In this section, we present two equivalent geometric constructions of all such quadratic 1D quasilattices.  Our formulation is designed to clarify the relationship between cut-and-project sequences, on the one hand, and lattice equivalence, self-similarity and self-sameness, on the other.  In the process, we obtain a number of explicit expressions that will be needed in subsequent sections, and in our construction of higher-dimensional Ammann patterns in \cite{BoyleSteinhardtHigherD}.

The starting point for both constructions is the same: an arbitrary Bravais lattice $\Lambda$ in 2D Euclidean space sliced by an arbitrary line $\vec{\;\!q}(t)$; and a choice of a ``positive" integer basis $\{\vec{m}_{1},\vec{m}_{2}\}$ for $\Lambda$.  We begin, then, by introducing these three ingredients.

\subsection{Geometric preliminaries: the lattice $\Lambda$, the line $\vec{\;\!q}(t)$, and the basis $\{\vec{m}_{1},\vec{m}_{2}\}$}
\label{q_and_Lambda}
  
Let $\Lambda$ be an arbitrary lattice in 2D Euclidean space, and let $\{\vec{m}_{1},\vec{m}_{2}\}$ be a (not necessarily orthonormal) integer basis for the lattice: every point in $\Lambda$ may be written as a unique integer linear combination of the vectors $\vec{m}_{1}$ and $\vec{m}_{2}$.  If we regard $\vec{m}_{1}$ and $\vec{m}_{2}$ as column vectors, the corresponding dual basis $\{\widetilde{m}^{1},\widetilde{m}^{2}\}$ consists of the row vectors $\widetilde{m}^{1}$ 
and $\widetilde{m}^{2}$ defined by the matrix equation
\begin{equation}
  \left[\begin{array}{c} 
  \widetilde{m}^{1} \\
  \widetilde{m}^{2}\end{array}\right]
  =[\;\vec{m}_{1}\;\;\vec{m}_{2}\;]^{-1}
  \qquad\Rightarrow\qquad
  \widetilde{m}^{i}\vec{m}_{j}=\delta^{i}_{\;j}.
\end{equation}
Let $\vec{\;\!q}(t)$ be an arbitrary line slicing through this space, and let $\{\hat{e}_{\parallel},\hat{e}_{\perp}\}$ be an orthonormal basis adapted to it: $\hat{e}_{\parallel}$ points along the line, $\hat{e}_{\perp}$ points perpendicular to it, and we write:
\begin{equation}
  \vec{\;\!q}(t)=\vec{\;\!q}_{0}+\hat{e}_{\parallel}t.
\end{equation}  
We will always assume that $\vec{\;\!q}(t)$ has irrational slope with respect to the $\{\vec{m}_{1},\vec{m}_{2}\}$ basis: {\it i.e.}\ $(\tilde{m}^{2}\hat{e}_{\parallel})/(\tilde{m}^{1}\hat{e}_{\parallel}$ is irrational.  It will be convenient to split $\vec{\;\!q}_{0}$, $\vec{m}_{1}$ and $\vec{m}_{2}$ into their $\hat{e}_{\parallel}$ and $\hat{e}_{\perp}$ components:
\begin{subequations}
  \begin{eqnarray}
    \vec{\;\!q}_{0\,}&=&\;q_{0}^{\,\parallel}\hat{e}_{\parallel}+\;q_{0}^{\,\perp}\hat{e}_{\perp}, \\
    \vec{m}_{1}&=&m_{1}^{\parallel}\hat{e}_{\parallel}+m_{1}^{\perp}\hat{e}_{\perp}. \\
    \vec{m}_{2}&=&m_{2}^{\parallel}\hat{e}_{\parallel}+m_{2}^{\perp}\hat{e}_{\perp}.
  \end{eqnarray}
\end{subequations}
In this paper, we will usually focus on the case where $\{\vec{m}_{1},\vec{m}_{2}\}$ is a ``positive basis," meaning that (for $i=1,2$) it satisfies the following conditions
\begin{subequations}
  \label{positive_basis}
  \begin{eqnarray}
    \label{positive_basis_a}
    \widetilde{m}^{i}\hat{e}_{\parallel}&>&0, \\
    \label{positive_basis_b}
    \vec{m}_{i}\!\;\!\cdot\!\;\!\hat{e}_{\parallel}&>&0.
  \end{eqnarray}
\end{subequations}
Note that, since $\vec{m}_{1}$ and $\vec{m}_{2}$ are not assumed to be orthogonal, conditions (\ref{positive_basis_a}) and (\ref{positive_basis_b}) are not redundant. 
Eq.~(\ref{positive_basis_a}) says that the vector $\hat{e}_{\parallel}$ lies in the ``first quadrant" with respect to the $\{\vec{m}_{1},\vec{m}_{2}\}$ basis ({\it i.e.}\ if we expand $\hat{e}_{\parallel}=\alpha_{1}\vec{m}_{1}+\alpha_{2}\vec{m}_{2}$ in the $\{\vec{m}_{1},\vec{m}_{2}\}$ basis, then the coordinates $\alpha_{1}$ and $\alpha_{2}$ are both positive); and Eq.~(\ref{positive_basis_b}) says that $\vec{m}_{1}$ and $\vec{m}_{2}$ both have positive projections onto $\hat{e}_{\parallel}$ ({\it i.e.}\ $m_{1}^{\parallel}$ and $m_{2}^{\parallel}$ are both positive).  

There is no loss of generality in assuming condition (\ref{positive_basis_a}) since it can always be achieved by flipping the sign of one or both of the basis vectors $\vec{m}_{1}$ and $\vec{m}_{2}$, as needed.  But there {\it is} the following loss of generality in assuming condition (\ref{positive_basis_b}).   This second condition only allows sequences of points $x_{n}$ that can be reached by walking along the line using steps of two different lengths, $L$ and $S$, that are {\it all in the same direction}.  By contrast, dropping condition (\ref{positive_basis_b}) would allow $\hat{m}_{1}\cdot\hat{e}_{\parallel}$ and $\vec{m}_{2}\cdot\hat{e}_{parallel}$ to have opposite signs,  yielding sequences of points $x_{n}$ that could be reached by taking steps of the same lengths but in two different directions ({\it e.g.}\ each step of length $L$ could be to the right, while each step of length $S$ could be to the left).  When such a sequence of points is read from left to right (as opposed to in the sequence with which the points were reached by the walker), it appears to be constructed of more than two different lengths.  
We wish to exclude this situation, since it does not correspond to a 1D tiling in the usual sense (since the tiles overlap), and cannot be used to construct standard higher-dimensional Ammann patterns or Penrose-like tilings, for the same reason.
\begin{figure}
  \begin{center}
    \includegraphics[width=3.0in]{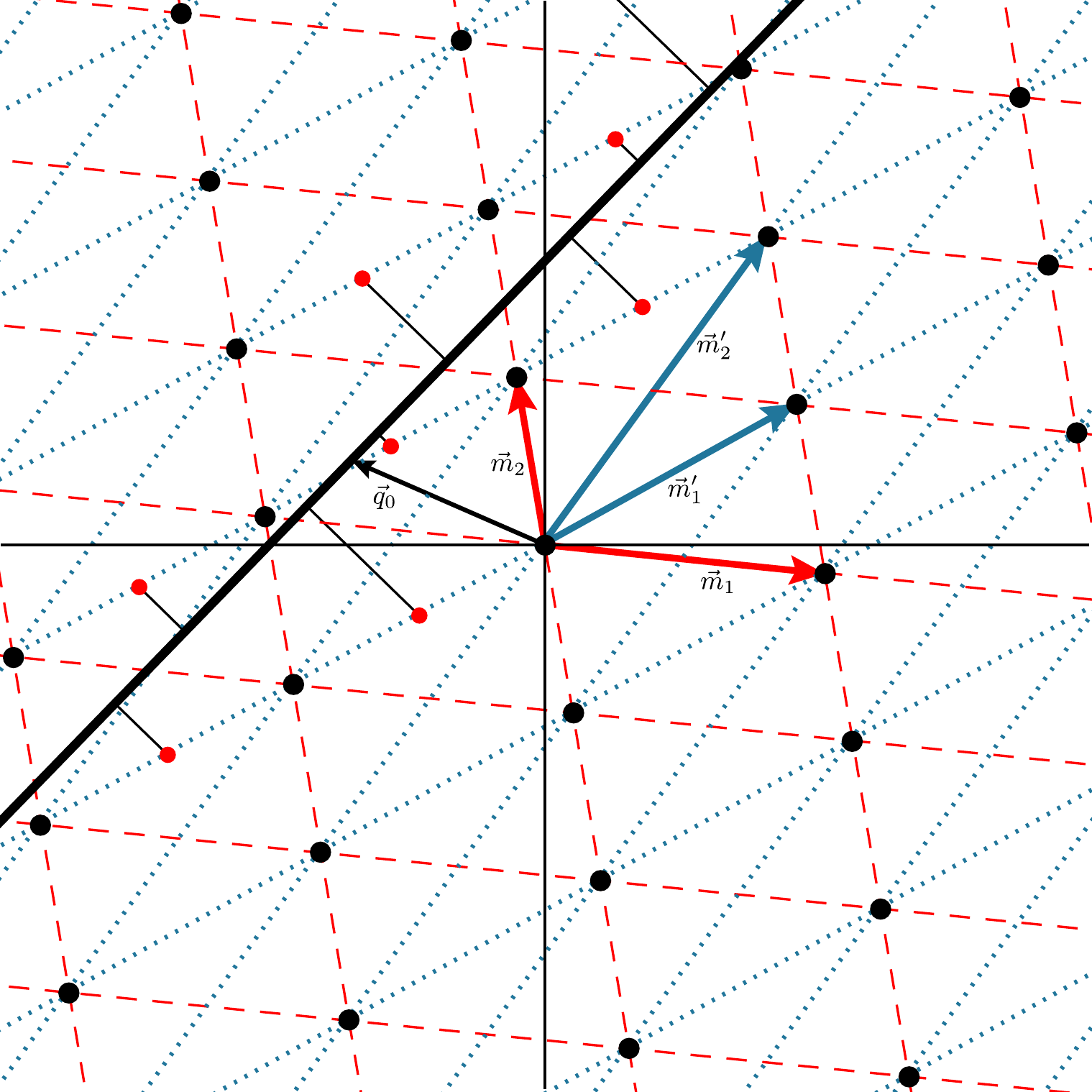}
  \end{center}
  \caption{Illustrates the geometric objects discussed in Sections II and III.  The black dots are the lattice $\Lambda$.  The thick black line is $\vec{\;\!q}(t)$, with its own origin displaced from the origin of $\Lambda$ by the vector $\vec{\;\!q}_{0}$.  The solid red arrows show an integer basis $\{\vec{m}_{1},\vec{m}_{2}\}$ for $\Lambda$, while the dashed red lines show the corresponding integer grid; and the figure illustrates the corresponding cut-and-project construction: every time the solid black line $\vec{\;\!q}(t)$ intersects one of the red dashed parallelograms, the midpoint of that paralleogram (a red dot) is orthogonally projected onto $\vec{\;\!q}(t)$ to obtain the 1D quasilattice $x_{n}$.  The solid turquoise arrows then show an alternative integer basis $\{\vec{m}_{1}',\vec{m}_{2}'\}$ for $\Lambda$, while the dotted turquoise lines show the corresponding integer grid; and this alternative basis could be used in an exactly analogous way to obtain a second quasilattice $x_{n}'$ which would be in the same equivalance class as the first: either one could be obtained from the other by a local decoration/gluing rule.}
 \label{CutAndProjectFig}
\end{figure}

\subsection{Perspective 1: the 1D quasilattice from dualizing a 1D bi-grid}
\label{dualization}

The $\{\vec{m}_{1},\vec{m}_{2}\}$ basis defines an ``integer grid": this is the set of all lines that (in the $\{\vec{m}_{1},\vec{m}_{2}\}$ basis) have a constant integer value for either their first or second coordinate (like the grid of lines on an ordinary sheet of graph paper).  The intersection of this integer grid with the line $\vec{\;\!q}(t)$ defines a 1D ``bi-grid."  In particular, the grid line whose first coordinate (in the $\{\vec{m}_{1},\vec{m}_{2}\}$ basis) is the integer $n\in\mathbb{Z}$ intersects $\vec{\;\!q}(t)$ at $t=t_{n}^{(1)}$, where
\begin{subequations}
  \begin{equation}
    \label{tn1}
    \widetilde{m}^{1}\vec{\;\!q}(t_{n}^{(1)})=n \qquad\Rightarrow\qquad t_{n}^{(1)}
    =\frac{n-\widetilde{m}^{1}\vec{\;\!q}_{0}}{\widetilde{m}^{1}\hat{e}_{\parallel}}, 
  \end{equation}
  while the grid line whose second coordinate (in the $\{\vec{m}_{1},\vec{m}_{2}\}$ basis) is the integer $n\in\mathbb{Z}$ intersects $\vec{\;\!q}(t)$ at $t=t_{n}^{(2)}$, where
  \begin{equation}
    \label{tn2}
    \widetilde{m}^{2}\vec{\;\!q}(t_{n}^{(2)})=n \qquad\Rightarrow\qquad t_{n}^{(2)}
    =\frac{n-\widetilde{m}^{2}\vec{\;\!q}_{0}}{\widetilde{m}^{2}\hat{e}_{\parallel}}. 
  \end{equation}
\end{subequations}
The points $t_{n}^{(1)}$ form a periodic 1D lattice of spacing $1/(\widetilde{m}^{1}\hat{e}_{\parallel})$, while the points $t_{n}^{(2)}$ form another periodic 1D lattice of spacing $1/(\widetilde{m}^{2}\hat{e}_{\parallel})$.  The superposition of these two periodic lattices (with incommensurate spacings) is the 1D bi-grid.  

From this 1D bi-grid, we obtain the corresponding 1D quasilattice by a standard ``dualization" procedure \cite{deBruijn81b, Senechal}: to each space between two consecutive points in the bi-grid, we assign a point $x$ in the dual quasilattice, so that (i) whenever we cross a point $t_{n}^{(1)}$ in the bi-grid (from the $t_{n-1}^{(1)}$ side to the $t_{n+1}^{(1)}$ side), we correspondingly jump $x\to x+m_{1}^{\parallel}$ in the dual quasi-lattice, and (ii) whenever we cross a point $t_{n}^{(2)}$ in the bi-grid (from the $t_{n-1}^{(2)}$ side to the $t_{n+1}^{(2)}$ side), we correspondingly jump $x\to x+m_{2}^{\parallel}$ in the dual quasi-lattice.  Stated another way, as the bi-grid parameter $t$ continuously sweeps from $-\infty$ to $+\infty$, the quasilattice point $x$ changes discretely, according to 
\begin{equation}
  \label{dualization_formula}
  x=\left\lfloor\widetilde{m}^{1}\vec{\;\!q}(t)\right\rfloor m_{1}^{\parallel}+\left\lfloor\widetilde{m}^{2}\vec{\;\!q}(t)\right\rfloor m_{2}^{\parallel}+C
\end{equation}
where $C$ is a constant.  

We can canonically fix $C$ by demanding that the quasilattice dual to a reflection-symmetric bi-grid is also reflection-symmetric, which fixes $C$ to be:
\begin{equation}
  \label{C}
  C=\frac{1}{2}m_{1}^{\parallel}+\frac{1}{2}m_{2}^{\parallel}-q_{0}^{\parallel}.
\end{equation}
Fixing this phase relationship is unimportant in 1D, but plays an important role when we construct higher-dimensional Ammann patterns in \cite{BoyleSteinhardtHigherD}, since these higher-dimensional Ammann patterns are built from a collection of multiple 1D quasilattices whose phases must be carefully coordinated with one another.  

\subsection{Perspective 2: the 1D quasilattice from a cut-and-project algorithm}
\label{cut_and_project}

Eqs.~(\ref{dualization_formula}) and (\ref{C}) also have another geometric interpretation.  We can think of the $\{\vec{m}_{1},\vec{m}_{2}\}$ integer grid described in Subsection \ref{dualization} as slicing up the plane into parallelgrams whose edges are the vectors $\vec{m}_{1}$ and $\vec{m}_{2}$, and whose vertices coincide with the points of $\Lambda$.  Now we can construct our 1D quasilattice by the following ``cut-and-project" algorithm: whenever the ``cut" line $\vec{\;\!q}(t)$ intersects one of these parallelograms, we orthogonally project the midpoint of that parallelogram onto the cut line to obtain the point $\vec{\;\!q}(x)$ (see Fig. 1).  This mapping from $t$ to $x$ is precisely the one described by Eqs.~(\ref{dualization_formula}) and (\ref{C}). In particular, fixing $C$ according to (\ref{C}) corresponds to projecting the parallelogram's midpoint.  This makes sense: by projecting the midpoint (as opposed to say the upper right corner) of each intersected parallelogram, we make the algorithm explicitly reflection-symmetric, so it must yield the same constant $C$ (\ref{C}) that that we obtained by demanding reflection symmetry in the previous (Perspective 1) algorithm.  (And this requirement that the algorithm respects reflection symmetry is also the reason that the substitution rules derived later are always reflection symmetric.)

This cut-and-project perspective leads to a convenient way of re-expressing Eqs.~(\ref{dualization_formula}) and (\ref{C}).  For the rest of this section, let us assume that $\{\vec{m}_{1},\vec{m}_{2}\}$ are a {\it positive} basis (see Subsection \ref{q_and_Lambda}).  As $t$ runs from $-\infty$ to $+\infty$, the line $\vec{\;\!q}(t)$ passes from one parallelogram to the next, thereby placing the parallelograms that it intersects in a specific order, which can be indexed by the integer $n$.  In particular, when $\vec{\;\!q}(t)$ passes through the $n$th parallelogram, it intersects that parallelogram's transverse diagonal at a time $t_{n}$ given by:
\begin{equation}
  \widetilde{m}\vec{\;\!q}(t_{n})=n\qquad\Rightarrow\qquad
  t_{n}=\frac{n-\widetilde{m}\vec{\;\!q}_{0}}{\widetilde{m}\hat{e}_{\parallel}}
\end{equation}
where 
\begin{equation}
  \widetilde{m}=\widetilde{m}^{1}+\widetilde{m}^{2}.
\end{equation}
We can use Eqs.~(\ref{dualization_formula}) and (\ref{C}) to map this $n$th intersection time, $t_{n}$, to a corresponding $n$th point in the quasilattice, $x_{n}$.  Following this procedure and massaging the result a bit, we obtain the useful formula:
\begin{eqnarray}
  \label{xn_symmetric}
  x_{n}
  &=&\left(\left\lfloor\frac{n m_{2}^{\perp}-q_{0}^{\perp}}{m_{2}^{\perp}-m_{1}^{\perp}}\right\rfloor+\frac{1}{2}\right)m_{1}^{\parallel}\nonumber\\
  &+&\left(\left\lfloor\frac{n m_{1}^{\perp}-q_{0}^{\perp}}{m_{1}^{\perp}-m_{2}^{\perp}}\right\rfloor+\frac{1}{2}\right)m_{2}^{\parallel}-q_{0}^{\parallel}.
\end{eqnarray}
Note that the $1/2$'s in this expression arise because we project the {\it midpoint} of each parallelogram (as opposed to one of its corners, say).  Let us make three remarks about Eq.~(\ref{xn_symmetric}):
\begin{enumerate}

\item Eq.~(\ref{xn_symmetric}) defines the same 1D quasilattice as Eqs.~(\ref{dualization_formula}, \ref{C}); the difference is that, whereas (\ref{dualization_formula}, \ref{C}) expressed this quasilattice as the range of a many-to-one map with a continuous domain ($t\in\mathbb{R}$), (\ref{xn_symmetric}) expresses the same quasilattice as the range of a one-to-one map from a discrete domain ($n\in\mathbb{Z}$).

\item Imagine replacing the line $\vec{\;\!q}=\vec{\;\!q}_{0}+\hat{e}_{\parallel}t$ by a new line $\vec{\;\!q}\,'=\vec{\;\!q}_{0}\!\!'+\hat{e}_{\parallel}t$ which is parallel to the original line, and has just been translated by a vector in $\Lambda$: $\vec{\;\!q}_{0}\!\!'=\vec{\;\!q}_{0}+n_{1}\vec{m}_{1}+n_{2}\vec{m}_{2}$ (for some integers $n_{1}$ and $n_{2}$).  Then, via Eq.~(\ref{xn_symmetric}), we obtain a new quasilattice $x_{n}'$ with correspondingly shifted parameters:
\begin{equation}
  \label{umklaap_symmetric}
  \begin{array}{c}
  q_{0}^{\,\parallel\;\!}{}'=q_{0}^{\,\parallel\;\!}+n_{1}m_{1}^{\,\parallel\,}+n_{2}m_{2}^{\,\parallel\,}, \\
  q_{0}^{\perp}{}'=q_{0}^{\perp}+n_{1}m_{1}^{\perp}+n_{2}m_{2}^{\perp}.
  \end{array}
\end{equation}
But, as may be checked from Eq.~(\ref{xn_symmetric}), the two quasilattices $x_{n}$ and $x_{n}'$ are actually identical up to reindexing: $x_{n}'=x_{n-n_{1}-n_{2}}$.  This is called an {\it umklaap transformation} \cite{SocolarSteinhardt86} and reflects the fact that, when we consider the family of 1D quasilattices obtained by varying $\vec{\;\!q}_{0}$, we can really think of $\vec{\;\!q}_{0}$ as living on a torus \cite{BaakeHermissonPleasants, HermissonRichardBaake}.

\item Since $\{\vec{m}_{1},\vec{m}_{2}\}$ is a positive basis, Eq.~(\ref{positive_basis_a}) requires $m_{2}^{\perp}/m_{1}^{\perp}<0$, and Eq.~(\ref{positive_basis_b}) requires $m_{1}^{\parallel}>0$ and $m_{2}^{\parallel}>0$.  Together these conditions imply that, as the integer index $n$ increments (from $n'$ to $n'+1$), the corresponding quasilattice position $x_{n}$ (\ref{xn_symmetric}) increases by one of the two positive lengths: $m_{1}^{\parallel}$ or $m_{2}^{\parallel}$. Furthermore,
\begin{equation}
  (f_{1}/f_{2})=-(m_{2}^{\perp}/m_{1}^{\perp}),
\end{equation}
where $f_{1}/f_{2}$ is the relative frequency of steps of length $m_{1}^{\parallel}$ and steps of length $m_{2}^{\parallel}$.  If $f_{1}/f_{2}<1$, the quasilattice consists of single (isolated) steps of length $m_{1}^{\parallel}$, separated by either $\lfloor f_{2}/f_{1}\rfloor$ or ($\lfloor f_{2}/f_{1}\rfloor+1$) steps of length $m_{2}^{\parallel}$; and if $f_{2}/f_{1}<1$, the quasilattice consists of single (isolated) steps of length $m_{2}^{\parallel}$, separated by either $\lfloor f_{1}/f_{2}\rfloor$ or ($\lfloor f_{1}/f_{2}\rfloor+1$) steps of length $m_{1}^{\parallel}$.

\end{enumerate}

Although Eq.~(\ref{xn_symmetric}) has the advantage of being manifestly symmetric under interchange of $1\leftrightarrow2$ subscripts, it is sometimes convenient to rewrite it in one of the following two forms, which each only involve one floor function $\lfloor\ldots\rfloor$, and are swapped by swapping $1\leftrightarrow2$:
\begin{subequations}  
  \label{xn_asymmetric} 
  \begin{eqnarray}
    \label{xn_asymmetric1}
    x_{n}
    \!&\!=\!&\!m_{1}^{\parallel}(n\!-\!\chi_{1}^{\parallel})\!+\!(m_{2}^{\parallel}\!-\!m_{1}^{\parallel})\!\left(\!\left\lfloor\kappa_{1}(n\!-\!\chi_{1}^{\perp})\right\rfloor
    \!+\!\frac{1}{2}\!\right) \\
    \label{xn_asymmetric2}
    \!&\!=\!&\!m_{2}^{\parallel}(n\!-\!\chi_{2}^{\parallel})\!+\!(m_{1}^{\parallel}\!-\!m_{2}^{\parallel})\!\left(\!\left\lfloor\kappa_{2}(n\!-\!\chi_{2}^{\perp})\right\rfloor
    \!+\!\frac{1}{2}\!\right)\qquad
  \end{eqnarray}
\end{subequations}
where we have defined the constants
\begin{subequations}
  \begin{eqnarray}
    \chi_{1}^{\parallel}\!\equiv\!q_{0}^{\parallel}/m_{1}^{\parallel},
    &\quad\!
    \chi_{1}^{\perp}\!\equiv\!q_{0}^{\perp}/m_{1}^{\perp}, 
    &\quad\! 
    \kappa_{1}\!\equiv\!\frac{m_{1}^{\perp}}{m_{1}^{\perp}\!-\!m_{2}^{\perp}}, \\
    \chi_{2}^{\parallel}\!\equiv\!q_{0}^{\parallel}/m_{2}^{\parallel},
    &\quad\!
    \chi_{2}^{\perp}\!\equiv\! q_{0}^{\perp}/m_{2}^{\perp}, 
    &\quad\! 
    \kappa_{2}\!\equiv\!\frac{m_{2}^{\perp}}{m_{2}^{\perp}\!-\!m_{1}^{\perp}}.\qquad
  \end{eqnarray}
\end{subequations}
Note that Eqs. (\ref{xn_asymmetric1}, \ref{xn_asymmetric2}) have the same form as our original "floor form" expression (\ref{second_order}), except that we have now switched notation (from $L$ and $S$ to $m_{1}^{\parallel}$ and $m_{2}^{\parallel}$ for the two tile lengths, and from $\alpha$ and $\beta$ to $\chi^{\parallel}$ and $\chi^{\perp}$ for the two phases) to specify more precisely the relationship to the underlying 2D lattice.  We add three more remarks:
\begin{enumerate}

\item When we re-express the quasilattice (\ref{xn_symmetric}) in the form (\ref{xn_asymmetric}), we correspondingly re-express the umklaap transformation (\ref{umklaap_symmetric}) in the form
\begin{subequations}
  \label{umklaap_asymmetric}
  \begin{equation}
    \label{umklaap_asymmetric1}
    \begin{array}{lcl}
    \chi_{1}^{\,\parallel}{}'&=&\chi_{1}^{\,\parallel}+n_{1}+n_{2}(m_{2}^{\,\parallel}/m_{1}^{\,\parallel}), \\
    \chi_{1}^{\perp}{}'&=&\chi_{1}^{\perp}+n_{1}+n_{2}(m_{2}^{\perp}/m_{1}^{\perp}), 
    \end{array}
  \end{equation}
  \begin{equation}
    \label{umklaap_asymmetric2}
    \begin{array}{lcl}
    \chi_{2}^{\,\parallel}{}'&=&\chi_{2}^{\,\parallel}+n_{2}+n_{1}(m_{1}^{\,\parallel}/m_{2}^{\,\parallel}), \\
    \chi_{2}^{\perp}{}'&=&\chi_{2}^{\perp}+n_{2}+n_{1}(m_{1}^{\perp}/m_{2}^{\perp}).
    \end{array}
  \end{equation}
\end{subequations}  

\item In the generic (non-singular) case where the line $\vec{\;\!q}(t)$ does not intersect any of the points in $\Lambda$, the three expressions (\ref{xn_symmetric}, \ref{xn_asymmetric1}, \ref{xn_asymmetric2}) are all equivalent.  In the special (singular) case where the line $\vec{\;\!q}(t)$ intersects a point in $\Lambda$, the three expressions (\ref{xn_symmetric}, \ref{xn_asymmetric1}, \ref{xn_asymmetric2}) are {\it almost} equivalent, but they differ at one point $x_{n_{\ast}}$ (where the argument of the floor function $\lfloor\ldots\rfloor$ is precisely an integer).  This may seem like a minor detail, but in fact (as we shall explain in a subsequent paper \cite{BoyleSteinhardtDefects}) these special cases are not only the 1D analogues of, but also the 1D building blocks for, a fascinating set of topological defects which are intrinsic to two- and higher-dimensional Penrose-like tilings (and are known as ``decapod defects" in the case of the standard 2D Penrose tiling \cite{Gardner77, GrunbaumShephard}).  For this reason, we will continue to keep track of this detail at later points in this paper (see Section \ref{self_similar} and Appendix \ref{singular}).

\item Comparing Eqs.~(\ref{second_order}) and (\ref{xn_asymmetric}), we see that {\it given a line $\vec{\;\!q}(t)$, a lattice $\Lambda$, and a positive basis $\{\vec{m}_{1},\vec{m}_{2}\}$, the cut-and-project algorithm described above produces a quadratic 1D quasilattice $x_{n}$}.  Conversely, it is easy to check that {\it any quadratic 1D quasilattice $x_{n}$ may be obtained from such a cut-and-project algorithm}: the ``$\parallel$" components of $\vec{m}_{1}$ and $\vec{m}_{2}$ can be chosen to obtain the desired parameters $S$ and $L$, the ``$\perp$" components of  $\vec{m}_{1}$ and $\vec{m}_{2}$ can be chosen to obtain the desired $\kappa$, and the ``$\parallel$" and ``$\perp$" components of $\vec{\;\!q}_{0}$ can be chosen to obtain the desired $\{\alpha,\beta\}$.

\end{enumerate}

\section{Lattice-equivalent quasilattices}
\label{lattice_equivalent}

In Section \ref{
two_perspectives}, we constructed the quadratic 1D quasilattice $x_{n}$ (\ref{xn_symmetric}) by first choosing: (i) a line $\vec{\;\!q}(t)$, (ii) a lattice $\Lambda$, and (iii) a ``positive" integer basis $\{\vec{m}_{1},\vec{m}_{2}\}$ for $\Lambda$.  If, instead, we choose the {\it same} line $\vec{\;\!q}(t)$ and the {\it same} lattice $\Lambda$, but a {\it different} positive integer basis $\{\vec{m}_{1}',\vec{m}_{2}'\}$, we obtain a different quadratic quasilattice:
\begin{eqnarray}
  \label{xn_symmetric_prime}
  x_{n}'&=&\left(\left\lfloor\frac{n m_{2}^{\perp}{}'-q_{0}^{\perp}}{m_{2}^{\perp}{}'-m_{1}^{\perp}{}'}\right\rfloor+\frac{1}{2}\right)m_{1}^{\parallel}{}'\nonumber\\
  &+&\left(\left\lfloor\frac{n m_{1}^{\perp}{}'-q_{0}^{\perp}}{m_{1}^{\perp}{}'-m_{2}^{\perp}{}'}\right\rfloor+\frac{1}{2}\right)m_{2}^{\parallel}{}'-q_{0}^{\parallel}.
\end{eqnarray}
We will call two quasilattices $x_{n}$ and $x_{n}'$ that are related in this way ``lattice equivalent."

To understand lattice equivalence in more detail, let us write the relationship between the two bases as
\begin{subequations}
  \label{m_prime_from_m}
  \begin{eqnarray}
    \vec{m}_{1}'&=&a\vec{m}_{1}+b\vec{m}_{2}, \\
    \vec{m}_{2}'&=&c\vec{m}_{1}+d\vec{m}_{2}.
  \end{eqnarray}
\end{subequations}
Since $\{\vec{m}_{1},\vec{m}_{2}\}$, $\{\vec{m}_{1}',\vec{m}_{2}'\}$ are both integer bases for $\Lambda$,
\begin{equation}
  \tau=\left(\begin{array}{cc} a & b \\ c & d \end{array}\right)
\end{equation}
must be an integer matrix with determinant $\pm1$.  And since $\{\vec{m}_{1},\vec{m}_{2}\}$ and $\{\vec{m}_{1}',\vec{m}_{2}'\}$ are both {\it positive} integer bases for $\Lambda$ and, without loss of generality, we take the unprimed $\{\vec{m}_{1},\vec{m}_{2}\}$ parallelogram to be the one that is wider in the $\hat{e}_{\perp}$ direction ($|m_{2}^{\perp}-m_{1}^{\perp}|>|m_{2}^{\perp}{}'-m_{1}^{\perp}{}'|$), the components $\{a,b,c,d\}$ of $\tau$ must also all be non-negative (see Appendix \ref{positivity_appendix} for a proof).

\begin{figure}
  \begin{center}
    \includegraphics[width=3.0in]{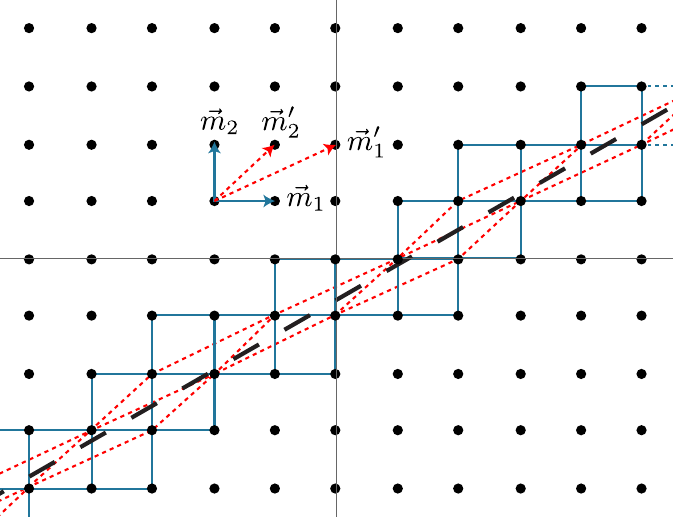}
  \end{center}
  \caption{Illustrates why two lattice equivalent quasilattices are related by a fixed decoration rule, as explained in Section \ref{lattice_equivalent}.}
 \label{LatticeEquivalentFig1}
\end{figure}
\begin{figure}
  \begin{center}
    \includegraphics[width=3.0in]{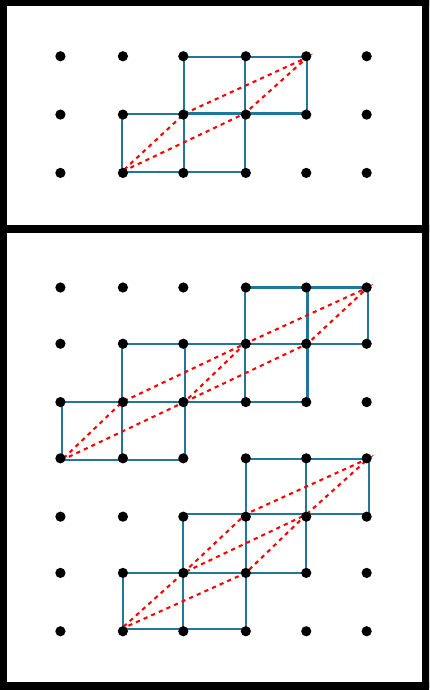}
  \end{center}
  \caption{More about the relationship between the parallelograms in Fig.~\ref{LatticeEquivalentFig1}: the top panel shows the minimal covering of one (red, elongated, dotted) $\{\vec{m}_{1}',\vec{m}_{2}'\}$ parallelogram by (blue, square, solid) $\{\vec{m}_{1},\vec{m}_{2}\}$ parallelograms; and the bottom panel shows the minimal covering of two adjacent $\{\vec{m}_{1}',\vec{m}_{2}'\}$, which just depends on whether they share a common short edge, or a common long edge.}
 \label{LatticeEquivalentFig2}
\end{figure}

The two lattice-equivalent quasilattices $x_{n}$ (\ref{xn_symmetric}) and $x_{n}'$ (\ref{xn_symmetric_prime}) are intimately related to one another: the denser quasilattice $x_{n}$ may be obtained from the sparser quasilattice $x_{n}'$ by applying a local ``substitution" or ``decoration" rule that replaces each type of interval between points in the sparser quasilattice by a specific, fixed sequence of intervals in the denser quasilattice; and in the other direction, the sparser quasilattice may be recovered from the denser one by a local rule for gluing together a certain specific, fixed sequence of intervals in the denser lattice to obtain each type of interval in the sparser one.  

To understand this assertion, consider Fig.~\ref{LatticeEquivalentFig1}: it shows the set of (red, elongated, dotted) $\{\vec{m}_{1}',\vec{m}_{2}'\}$ parallelograms and the set of (blue, square, solid) $\{\vec{m}_{1},\vec{m}_{2}\}$ parallelograms that are intersected by the diagonal black dotted line $\vec{\;\!q}(t)$.  Note that the set of $\{\vec{m}_{1},\vec{m}_{2}\}$ parallelograms is precisely the minimal set needed to cover the set of $\{\vec{m}_{1}',\vec{m}_{2}'\}$ parallelograms completely.  (For clarity, in the top panel of Fig.~\ref{LatticeEquivalentFig2} we show a single 
$\{\vec{m}_{1}',\vec{m}_{2}'\}$ parallelogram and its minimal covering by $\{\vec{m}_{1},\vec{m}_{2}\}$ parallelograms; and in the bottom panel of Fig.~\ref{LatticeEquivalentFig2}, we show the minimal covering of two adjacent $\{\vec{m}_{1}',\vec{m}_{2}'\}$ parallelograms, depending on whether they share a common long edge or a common short edge.)  From Figs.~\ref{LatticeEquivalentFig1} and \ref{LatticeEquivalentFig2}, we can see the simple geometric reason why (as asserted above) the denser quasilattice $x_{n}$ may be obtained from the sparser quasilattice $x_{n}'$ by applying a fixed ``substitution" or ``decoration" rule to each of the two intervals ($S'$ and $L'$) in the $x_{n}'$ quasilattice -- it is because: (i) whenever two adjacent $\{\vec{m}_{1}',\vec{m}_{2}'\}$ parallelograms share a common {\it short} edge (giving rise to an $L'$ interval in this example), they are always covered by the same arrangement of $\{\vec{m}_{1},\vec{m}_{2}\}$ parallelograms, which yields a fixed decoration of $L'$ by $S$ and $L$; and (ii) whenever two adjacent $\{\vec{m}_{1}',\vec{m}_{2}'\}$ parallelograms share a common {\it long} edge (giving rise to an $S'$ interval in this example), they are covered by the same arrangement of $\{\vec{m}_{1},\vec{m}_{2}\}$ parallelograms, yielding a fixed decoration of $S'$ by $S$ and $L$.  

Lattice equivalence thus organizes the various quadratic 1D quasilattices $x_{n}$ obtained the previous section into lattice equivalence classes (with an uncountable infinity of different lattice equivalence classes, and a countable infinity of different quasilattices in any particular lattice equivalence class).  Given a fixed line $\vec{\;\!q}(t)$ and a fixed lattice $\Lambda$, the various members of the corresponding lattice equivalence class come from all the different ways of choosing a positive integer basis $\{\vec{m}_{1},\vec{m}_{2}\}$ for $\Lambda$; and any two members of the family may be derived from one another by a local substitution/gluing rule corresponding to the integer matrix $\tau$.

The notion of lattice equivalence elucidates the precise connection between substitution sequences, on the one hand, and cut-and-project sequences, on the other. On the one hand, if the quasilattices $x_{n}$ and $x_{n}'$ are lattice equivalent, the sequence $x_{n}$ (regarded as an infinite string of unprimed letters $S$ and $L$) may be algebraically obtained from the sequence $x_{n}'$ (regarded as an infinite string of primed letters $S'$ and $L'$) by a formal substitution rule in which each primed letter ($S'$ or $L'$) is replaced by a fixed finite string of unprimed letters ($S$ and $L$).  This substitution rule may be summarized by a $2\times2$ integer matrix $\tau$ in a standard way.  On the other hand, we see that this {\it same} matrix $\tau$ has a simple geometric meaning: it is precisely the matrix that connects the unprimed basis $\{\vec{m}_{1},\vec{m}_{2}\}$ (that produces $x_{n}$ via cut-and-project) to the primed basis $\{\vec{m}_{1}',\vec{m}_{2}'\}$ (that produces $x_{n}'$ via cut-and-project).  

Furthermore, from the algebraic perspective, the matrix $\tau$ is not enough to specify the substitution rule, since it doesn't fix the particular ordering or overall translational phase of the substitution rule.  For example, 
\begin{equation}
  \label{example_tau}
  \tau=\left(\begin{array}{cc} 1 & 1 \\ 2 & 1 \end{array}\right)
\end{equation}
might correspond to any of the following rules:
\begin{subequations}
  \begin{eqnarray}
    \label{sub1}
    \{S',L'\}&\to&\{\frac{L}{2}S\frac{L}{2},\frac{L}{2}SS\frac{L}{2}\} \\
    \label{sub2}
    \{S',L'\}&\to&\{LS,LSS\} \\
    \label{sub3}
    \{S',L'\}&\to&\{LS,SLS\}. 
  \end{eqnarray}
\end{subequations}
Note that, (\ref{sub1}) and (\ref{sub2}) correspond to different phases, while (\ref{sub3}) corresponds to a different ordering -- {\it i.e.}\ if we start from the same parent sequence and apply these substitutions, then (\ref{sub1}) and (\ref{sub2}) will produce two sequences that only differ by an overall translation by $L/2$, while (\ref{sub2}) and (\ref{sub3}) will produce two genuinely distinct daughter sequences.  By contrast, from the geometrical perspective, the matrix $\tau$ {\it also} determines an ordering and a phase -- {\it i.e.} it is associated with a canonical substitution rule and, in particular, one that is $x\to-x$ reflection symmetric.  For example, for $\tau$ given by Eq.~(\ref{example_tau}), the canonical substitution rule is given by (\ref{sub1}) -- see Row 2a in Table I.  These canonical substitution rules (including ordering and phase) will be important in our analysis of higher-dimensional Ammann patterns in \cite{BoyleSteinhardtHigherD}.

\section{Self-similar quasilattices}
\label{self_similar}

\begin{figure}
  \begin{center}
    \includegraphics[width=3.0in]{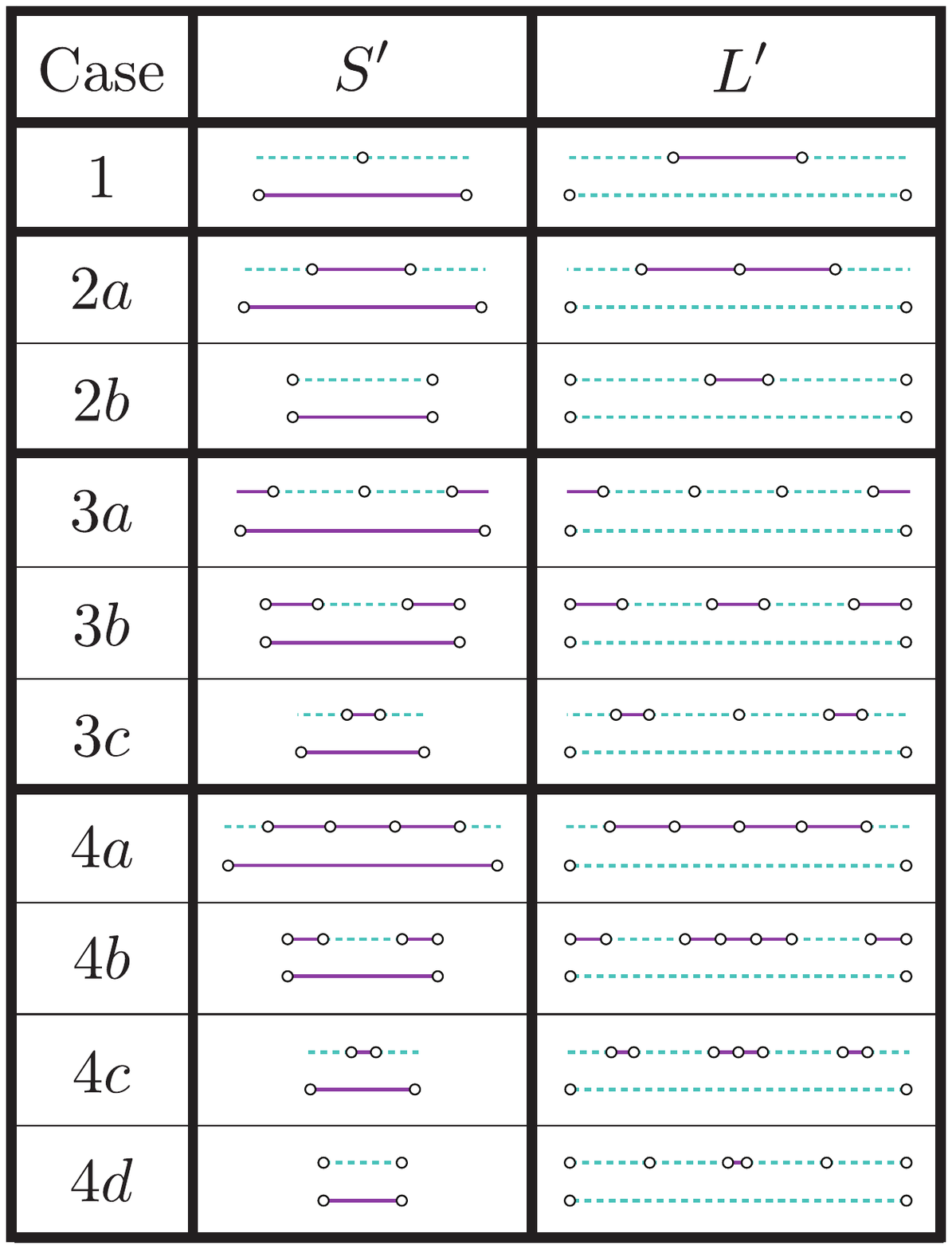}
  \end{center}
  \caption{Illustrations of the ten 1D self-similar substitution rules relevant to constructing higher-dimensional Ammann patterns and Penrose-like tilings (as catalogued in the last column of Table I).  In each row of this figure, the short (solid, purple) and long (dashed, turqoise) prototiles are on the bottom, with their corresponding self-similar decimations into smaller tiles directly above.  Open circles indicate the endpoints of tiles.  Complete tiles have have circles at both ends; half tiles have a circle at one end but none at the half-way point.  For example, Row 1 shows how a short prototile $S'$ (bottom left) is subdivided into two halves of a long prototile: $S'=(L/2)(L/2)$ (top left); and a long prototile $L'$ (bottom right) is subdivided into $L'=(L/2)S(L/2)$ (top right).}
 \label{1DSubstitutionFig}
\end{figure}

Two 1D quasilattices are locally isomorphic if any finite segment which occurs in one quasilattice also occurs somewhere in the other quasilattice, and vice versa, so that it is impossible, by inspecting any finite segment, to determine which of the two quasilattices one is looking at (see {\it e.g.}\ \cite{Gardner77, SocolarSteinhardt86, BaakeGrimme} for more).  Two lattice-equivalent quasilattices $x_{n}$ and $x_{n}'$, related by the matrix $\tau$, may look very different from one another and, in general, will not be locally isomorphic, even after an overall rescaling.   If $x_{n}$ and $x_{n}'$ are {\it also} locally isomophic (up to overall rescaling), then we say they are {\it self-similar} (under the transformation $\tau$).  In this section, we give the general closed-form expression for a self-similar quadratic 1D quasilattice, and a simple rule for how its parameters transform under a self-similarity (inflation/deflation) transformation.  Then, in Table I, we collect the ten special self-similar sequences that are relevant for constructing higher-dimensional Ammann patterns \cite{BoyleSteinhardtHigherD}; and in Figure \ref{1DSubstitutionFig}, we depict the corresponding substitution rules.

To start, let us pick a particular transformation matrix 
\begin{equation}
  \tau=\left(\begin{array}{cc} a & b \\ c & d \end{array}\right)
\end{equation}
(with non-negative integer components and determinant $\pm1$).  In the self-similar case, the new quasilattice $x_{n}'$ (\ref{xn_symmetric_prime}) is related to the original one $x_{n}$ (\ref{xn_symmetric}) by
\begin{equation}
  \label{ratios}
  \frac{m_{1}^{\parallel}{}'}{m_{2}^{\parallel}{}'}=
  \frac{m_{1}^{\parallel}}{m_{2}^{\parallel}}
  \qquad{\rm and}\qquad
  \frac{m_{1}^{\perp}{}'}{m_{2}^{\perp}{}'}=
  \frac{m_{1}^{\perp}}{m_{2}^{\perp}}; \end{equation}
  or, equivalently, 
\begin{equation}
  \left(\begin{array}{c} m_{1}^{\parallel} \\ m_{2}^{\parallel} \end{array}\right)
  \qquad{\rm and}\qquad
  \left(\begin{array}{c} m_{1}^{\perp} \\ m_{2}^{\perp} \end{array}\right)
\end{equation}
must  be  two {\it different} eigenvectors of $\tau$ with corresponding eigenvalues
\begin{subequations}
  \begin{eqnarray}
    \lambda_{\parallel}&=&\frac{1}{2}\left[a+d+\sqrt{(a+d)^{2}-4(ad-bc)}\right] \\
    \lambda_{\perp}&=&\frac{1}{2}\left[a+d-\sqrt{(a+d)^{2}-4(ad-bc)}\right]\qquad
  \end{eqnarray}
\end{subequations}
where $\lambda_{\parallel}>1$, while $|\lambda_{\perp}|<1$ and ${\rm sign}(\lambda_{\perp})={\rm det}(\tau)$.  

Note that the ratio $m_{1}^{\parallel}/m_{2}^{\parallel}$ determines the relative length of the two tiles in the quasilattice, while the ratio $m_{1}^{\perp}/m_{2}^{\perp}$ determines the relative frequencies of the two tiles.  From Eq.~(\ref{xn_asymmetric}) we see that, by holding these two ratios fixed under inflation, we ensure that the old and new quasilattices $x_{n}$ and $x_{n}'$ are in the same local isomorphism class.   That is, up to an overall rescaling, they only differ in their phases $\{\chi^{\parallel}, \chi^{\perp}\}$ vs $\{\chi^{\parallel}{}',\chi^{\perp}{}'\}$, which only determine which representatives of the local isomorphism class we are considering.

If the quasilattice $x_{n}$ (\ref{xn_symmetric}, \ref{xn_asymmetric1}, \ref{xn_asymmetric2}) is self-similar with respect to the transformation $\tau$, then after $s$ successive transformations, the resulting sequence $x_{n,s}^{}$ is:
\begin{subequations} 
  \label{xn_s}
  \begin{eqnarray}
    \label{xn_s_symmetric}
    \frac{x_{n,s}}{\lambda_{\parallel}^{s}}
    \!&\!=\!&\!\left(\left\lfloor\frac{n m_{2}^{\perp}-q_{0,s}^{\perp}}{m_{2}^{\perp}-m_{1}^{\perp}}\right\rfloor+\frac{1}{2}\right)m_{1}^{\parallel}\nonumber\\
    \!&\!+\!&\!\left(\left\lfloor\frac{n m_{1}^{\perp}-q_{0,s}^{\perp}}{m_{1}^{\perp}-m_{2}^{\perp}}\right\rfloor+\frac{1}{2}\right)m_{2}^{\parallel}-q_{0,s}^{\parallel} \\
    \label{xn_s_asymmetric1}
    \!&\!=\!&\!m_{1}^{\parallel}(n\!-\!\chi_{1,s}^{\parallel})\!+\!(m_{2}^{\parallel}\!-\!m_{1}^{\parallel})\!\left(\!\left\lfloor\!\kappa_{1}(n\!-\!\chi_{1,s}^{\perp})\!\right\rfloor\!+\!
    \frac{1}{2}\!\right) \\
    \label{xn_s_asymmetric2}
    \!&\!=\!&\!m_{2}^{\parallel}(n\!-\!\chi_{2,s}^{\parallel})\!+\!(m_{1}^{\parallel}\!-\!m_{2}^{\parallel})\!\left(\!\left\lfloor\!\kappa_{2}(n\!-\!\chi_{2,s}^{\perp})\!\right\rfloor\!+\!
    \frac{1}{2}\!\right)\qquad\;\;
  \end{eqnarray}
\end{subequations}
with new parameters $\{q_{0,s}^{\parallel}, q_{0,s}^{\perp}\}$ (or $\{\chi_{1,s}^{\parallel}, \chi_{1,s}^{\perp}\}$ or $\{\chi_{2,s}^{\parallel}, \chi_{2,s}^{\perp}\}$) which are related to the original parameters  $\{q_{0}^{\parallel}, q_{0}^{\perp}\}$ (or $\{\chi_{1}^{\parallel}, \chi_{1}^{\perp}\}$ or $\{\chi_{2}^{\parallel}, \chi_{2}^{\perp}\}$) as follows
\begin{subequations}
  \label{inflation}
  \begin{eqnarray}
    q_{0,s}^{\parallel}\;\!\equiv\frac{q_{0}^{\parallel}}{\lambda_{\parallel}^{s}},
    &\qquad&
    \;\!q_{0,s}^{\perp}\;\!\equiv\frac{q_{0}^{\perp}}{\lambda_{\perp}^{s}}, \\
    \chi_{1,s}^{\parallel}\equiv\frac{\chi_{1}^{\parallel}}{\lambda_{\parallel}^{s}},
    &\qquad&
    \chi_{1,s}^{\perp}\equiv\frac{\chi_{1}^{\perp}}{\lambda_{\perp}^{s}}, \\
    \chi_{2,s}^{\parallel}\equiv\frac{\chi_{2}^{\parallel}}{\lambda_{\parallel}^{s}},
    &\qquad&
    \chi_{2,s}^{\perp}\equiv\frac{\chi_{2}^{\perp}}{\lambda_{\perp}^{s}}.
  \end{eqnarray}
\end{subequations}
The formulae for a single inflation transformation are obtained by substituting $s=1$ in Eqs.~(\ref{xn_s}, \ref{inflation}).

If we want to describe 1D quasilattices and their self-similarity transformations in a way that continues to be precisely correct, even in the singular case where $\vec{\;\!q}(t)$ intersects a point in $\Lambda$, we have to replace Eqs.~(\ref{xn_s_symmetric}, \ref{xn_s_asymmetric1}, \ref{xn_s_asymmetric2}) by Eqs.~(\ref{xn_s_symmetric}$'$, \ref{xn_s_asymmetric1}$'$, \ref{xn_s_asymmetric2}$'$) presented in Appendix \ref{singular}.  This case will be studied further in a subsequent paper 
 \cite{BoyleSteinhardtDefects} where these singular 1D quasilattices are related to the intrinsic defects which can arise in Penrose-like tilings (like the ``decapod" defects in the Penrose tiling \cite{Gardner77, GrunbaumShephard}).  

\begin{table}
\begin{center}
  \begin{tabular}{|l|c|c|c|c|c|}
    
    \hline
    
    Case & $\lambda_{\pm}$  & $\tau$ &  $m_{2}^{\pm}/m_{1}^{\pm}$ & $S'$ & $L'$ \\
    
    \hline\hline
    
    1 & $\frac{1}{2}(1\pm\sqrt{5})$ & 
    $\left(\begin{array}{cc} 0 & 1 \\ 1 & 1 \end{array}\right)$ & 
    $\frac{1}{2}(1\pm\sqrt{5})$ & $\frac{L}{2}\frac{L}{2}$ & $\frac{L}{2}S\frac{L}{2}$ \\
    
    \hline\hline
    
    2a & $1\pm\sqrt{2}$ & 
    $\left(\begin{array}{cc} 1 & 1 \\ 2 & 1 \end{array}\right)$ &
    $\pm\sqrt{2}$ & $\frac{L}{2}S\frac{L}{2}$ & $\frac{L}{2}SS\frac{L}{2}$ \\
    
    \hline
    
    2b & $1\pm\sqrt{2}$ & 
    $\left(\begin{array}{cc} 0 & 1 \\ 1 & 2 \end{array}\right)$ & 
    $1\pm\sqrt{2}$ & $L$ & $LSL$ \\
  
    \hline\hline
    
    3a & $2\pm\sqrt{3}$ & 
    $\left(\begin{array}{cc} 1 & 2 \\ 1 & 3 \end{array}\right)$ &
    $\frac{1}{2}(1\pm\sqrt{3})$ & $\frac{S}{2}LL\frac{S}{2}$ & $\frac{S}{2}LLL\frac{S}{2}$ \\
    
    \hline

    3b & $2\pm\sqrt{3}$ & 
    $\left(\begin{array}{cc} 2 & 1 \\ 3 & 2 \end{array}\right)$ & 
    $\pm\sqrt{3}$ & $SLS$ & $SLSLS$ \\
    
    \hline

    3c & $2\pm\sqrt{3}$ & 
    $\left(\begin{array}{cc} 1 & 1 \\ 2 & 3 \end{array}\right)$ &
    $1\pm\sqrt{3}$ & $\frac{L}{2}S\frac{L}{2}$ & $\frac{L}{2}SLLS\frac{L}{2}$ \\
    
    \hline\hline
    
    4a & $2\pm\sqrt{5}$ & 
    $\left(\begin{array}{cc} 3 & 1 \\ 4 & 1 \end{array}\right)$ & 
    $-1\pm\sqrt{5}$ & $\frac{L}{2}SSS\frac{L}{2}$ & $\frac{L}{2}SSSS\frac{L}{2}$ \\
    
    \hline
    
    4b & $2\pm\sqrt{5}$ & 
    $\left(\begin{array}{cc} 2 & 1 \\ 5 & 2 \end{array}\right)$ & 
    $0\pm\sqrt{5}$ & $SLS$ & $SLSSSLS$ \\
    
    \hline
    
    4c & $2\pm\sqrt{5}$ & 
    $\left(\begin{array}{cc} 1 & 1 \\ 4 & 3 \end{array}\right)$ &
    $1\pm\sqrt{5}$ & $\frac{L}{2}S\frac{L}{2}$ & $\frac{L}{2}SLSSLS\frac{L}{2}$ \\
    
    \hline
    
    4d & $2\pm\sqrt{5}$ & 
    $\left(\begin{array}{cc} 0 & 1 \\ 1 & 4 \end{array}\right)$ &
    $2\pm\sqrt{5}$ & $L$ & $LLSLL$ \\
    
    \hline
  
  \end{tabular}
\end{center}
\caption{Catalog of the ten 1D self-similar quasilattices relevant to constructing higher-dimensional Ammann patterns and Penrose-like tilings in \cite{BoyleSteinhardtHigherD}.  In this table, we use the convenient notation $\lambda_{\pm}$ and $m_{i}^{\pm}$ where here the superscript/subscript ``$+$" stands for the former subscript/superscript ``$\parallel$", while the ``$-$" stands for ``$\perp$".  Within each case, the subcases are in order of increasing $L/S$.}
\end{table}

In our subsequent paper \cite{BoyleSteinhardtHigherD}, where these self-similar 1D quasilattices are used as the building blocks for higher dimensional Ammann patterns and Penrose-like tilings, four cases are relevant (see Table I in \cite{ReflectionQuasilattices}): Case 1, where the scaling factor is the ``golden ratio", $\lambda_{\parallel}=\phi=(1+\sqrt{5})/2$, which is the relevant case for describing systems with 5-fold or 10-fold order in 2D, some systems with icosahedral ($H_{3}$) order in 3D, and systems with ``hyper-icosahedral" ($H_{4}$) order in 4D; Case 2, where the scale factor is the ``silver ratio" $\lambda_{\parallel}=(1+\sqrt{2})$, which is the relevant case for describing systems with 8-fold order in 2D; Case 3, where the scale factor is $\lambda_{\parallel}=(2+\sqrt{3})$, which is the relevant case for describing systems with 12-fold order in 2D; and Case 4, where the scale factor is $\lambda_{\parallel}=\phi^{3}=2+\sqrt{5}$, which is the relevant case for describing some systems with icosahedral ($H_{3}$) order in 3D.  In Table I we list all ten of the 1D self-similar quasilattices corresponding to these four cases, and provide the relevant parameters needed to describe them explicitly.\footnote{Note that some of these were already fully or partially described in the literature, and others seem not to have been.  Case 1 (the Fibonacci lattice) and its canonical substitution rule were known in one form or another at least since Ammann came across them in the 1970's, and an early instance of its explicit floor-form expression and the corresponding canonical inflation/deflation rule may be found in \cite{SocolarSteinhardt86}.  Case 2a was the subject of \cite{deBruijn81a} and is also in \cite{Socolar89}, along with Case 3a.  In addition, all three of the $2\times2$ transformation matrices corresponding to the ratio $(2+\sqrt{3})$ may be found in \cite{BaakeGrimme} (but not the corresponding closed-form expressions for the quasilattice and its self-similarity transformation, or the canonical substitution rule, including ordering and phase).}  Note that in this table we have used the convenient notation $\lambda_{\pm}$ and $m_{i}^{\pm}$ where here the ``$+$" superscript/subscript stands for the former superscript/subscript ``$\parallel$", and the ``$-$" stands for ``$\perp$".

Finally, it may be possible to construct 1D quasilattices that only return to the same local isomorphism class (up to rescaling) after $s$ iterations of the same substitution (inflation) rule.  These would be 1D analogues of the 2D ``Ammann cycles" discussed in our subsequent paper \cite{BoyleSteinhardtHigherD}.   We leave the study of these objects in 1D to future work, and thank the anonymous referee for pointing out this interesting possibility. 

\section{Self-same quasilattices}

In the previous section, we restricted our attention to 1D quasilattices that were {\it self-similar} ({\it i.e.}\ both lattice equivalent and locally isomorphic) under the transformation $\tau$.  In this section, we restrict our attention further to 1D quasilattices $x_{n}$ that are {\it $s$-fold self-same} with respect to $\tau$ -- meaning that $x_{n}$ is self-similar with respect to $\tau$ and, moreover, $x_{n,s}$ (the quasilattice obtained by performing $s$ successive $\tau$-transformations) is {\it identical} to $x_{n}$ (after an appropriate rescaling). 

In the previous section, we found that after $s$ successive $\tau$ transformations, the original quasilattice $x_{n}$ (\ref{xn_symmetric}, \ref{xn_asymmetric1}, \ref{xn_asymmetric2}) characterized by parameters $\{q_{0}^{\parallel}, q_{0}^{\perp}\}$ (or $\{\chi_{1}^{\parallel}, \chi_{1}^{\perp}\}$ or $\{\chi_{2}^{\parallel}, \chi_{2}^{\perp}\}$) was transformed to a new quasilattice $x_{n,s}$ (\ref{xn_s_symmetric}, \ref{xn_s_asymmetric1}, \ref{xn_s_asymmetric2}) characterized by new parameters $\{q_{0,s}^{\parallel}, q_{0,s}^{\perp}\}$ (or $\{\chi_{1,s}^{\parallel}, \chi_{1,s}^{\perp}\}$ or $\{\chi_{2,s}^{\parallel}, \chi_{2,s}^{\perp}\}$).  In order for $x_{n}$ to be $s$-fold self-same, these transformed parameters must be related to the original parameters by an umklaap transformation (\ref{umklaap_symmetric}, \ref{umklaap_asymmetric1}, \ref{umklaap_asymmetric2}).  This implies that a quasilattice will be $s$-fold self-same with respect to $\tau$ if it is self-similar with respect to $\tau$ and, in addition, its parameters are given by
\begin{subequations}
  \label{self_same}
  \begin{eqnarray}
    q_{0}^{\pm}=\frac{\lambda_{\pm}^{s}(n_{1}m_{1}^{\pm}+n_{2}m_{2}^{\pm})}{1-\lambda_{\pm}^{s}}, \\
    \chi_{1}^{\pm}=\frac{\lambda_{\pm}^{s}(n_{1}m_{1}^{\pm}+n_{2}m_{2}^{\pm})}{(1-\lambda_{\pm}^{s})m_{1}^{\pm}}, \\
    \chi_{2}^{\pm}=\frac{\lambda_{\pm}^{s}(n_{2}m_{2}^{\pm}+n_{1}m_{1}^{\pm})}{(1-\lambda_{\pm}^{s})m_{2}^{\pm}},
  \end{eqnarray}
\end{subequations}
for any integers $n_{1}$ and $n_{2}$ (where, again in this section, we are using the notation that superscripts/subscripts $+$ and $-$ stand for $\parallel$ and $\perp$, respectively).  

As it stands, this answer is redundant, because there can be different ordered pairs  $\{n_{1},n_{2}\}$ and $\{n_{1}',n_{2}'\}$ of integers for which the above parameters secretly describe the same quasilattice (up to umklaap).   In order to count the non-redundant self-same crystals, first note that, comparing Eqs.~(\ref{self_same}) to Eqs.~(\ref{umklaap_symmetric}, \ref{umklaap_asymmetric}), the umklaap equivalent values of $\vec{q}_{0}$ form a 2D lattice (whose fundamental domain is a parallelogram with edges $\vec{m}_{1}$ and $\vec{m}_{2}$) and the $s$-fold self-same lattices also form a 2D lattice but (relative to the fundamental domain of the umklaap lattice) its fundamental domain is rescaled by $\lambda_{\pm}^{s}/(1-\lambda_{\pm}^{s})$ along the $\hat{e}_{\pm}$ directions, respectively. 

The naive formula for the number of distinct $s$-fold self-same lattices is then the ratio of the area of these two fundamental domains: $|(1-\lambda_{+}^{s})(1-\lambda_{-}^{s})/(\lambda_{+}^{s}\lambda_{-}^{s})|$.  This answer is almost correct, but requires the following correction.  The $n_{1}=n_{2}=0$ quasilattice is singular, and is actually a pair of quasilattices that are nearly identical to one another: see Appendix \ref{singular}.  The pair of quasilattices only differ at the very middle, where one lattice has the sequence ``$LS$" while the other has ``$SL$".  Beyond this middle pair of intervals, the two quasilattices are reflection symmetric and identical to one another.  When ${\rm det}\,\tau=+1$, these two sequences each inflate into themselves ({\it i.e.}\ they are both $1$-fold self-same); and when ${\rm det}\,\tau=-1$, they inflate into each other ({\it i.e.}\ they are $2$-fold self-same).  Taking this correction into account, we find that the number of quasilattices that are $s$-fold self-same with respect to $\tau$ is given by
\begin{equation}
  N_{s}=\left|\frac{(1-\lambda_{+}^{s})(1-\lambda_{-}^{s})}{\lambda_{+}^{s}\lambda_{-}^{s}}\right|+({\rm det}\,\tau)^{s}.
\end{equation}
But this result is not yet what we want, since it includes quasilattices that are self-same after $s$ inflations, but were already self same after $r$ inflations, where $r$ is a divisor of $s$.  After we remove these ``reducible" cases, we are left with the number $N_{s}'$ of {\it irreducible} $s$-fold self-same quasilattices.  The value of $N_{s}'$ may be determined iteratively by the formula
\begin{equation}
  N_{s}'=N_{s}-\sum_{\substack{r<s \\ r|s}}N_{r}'.
\end{equation}
The number $N_{s}'$ is divisible by $s$, since the irreducible $s$-fold self-same quasilattices are grouped into families of size $s$ which cycle into one another under $\tau$-transformation, and which we will call ``$s$-cycles."  So the most natural thing to count is the number of $s$-cycles, $N_{s}'/s$.  in Table II, we tabulate the number of $s$-cycles for the four important scale factors catalogued in Table I. Note that self-same quasilattices are examples of fixed points in the torus parameterization; in this context, the number of irreducible $s$-cycles in the golden ratio case (Case 1) was previously computed in \cite{BaakeHermissonPleasants, HermissonRichardBaake}.
\begin{table}
\begin{center}
  \begin{tabular}{c||c|c|c|c}
    \hline
    $s$ & $N_{s}'/s$ & $N_{s}'/s$ & $N_{s}'/s$ & $N_{s}'/s$ \\
    \hline
    1 &  0 & 1 & 3 & 3 \\
    2 &  1 & 2 & 5 & 7 \\
    3 &  1 & 4 & 16 & 24  \\
    4 &  1 & 7 & 45 & 76 \\
    5 &  2 & 16& 144 & 272 \\
    6 &  2 & 30 & 440 & 948 \\
    7 &  4 & 68 & 1440 & 3496 \\
    8 &  5 & 140 & 4680 & 12920 \\
  \end{tabular}
\end{center}
\caption{Here we list tabulate the first 8 terms in the sequence $N_{s}'/s$, for the four scale factors in Table I: $\phi=(1+\sqrt{5})/2$ (column 1); $(1+\sqrt{2})$ (column 2); $(2+\sqrt{3})$ (column 3); and $(2+\sqrt{5})$ (column 4).}
\end{table}  

It is interesting to note that the sequences of numbers in some of the columns in Table II  already appear as entries in the Online Encyclopedia of Integer Sequences (OEIS) for various different reasons.  Here we mention those entries for completeness and in the hope that, by tracking down the relationships, some interesting insights might be uncovered.  The first column is A006206 (``Number of aperiodic binary necklaces of  length $n$ with no subsequence $00$, excluding the necklace $0$"); the second column is A215335 (``Cyclically smooth Lyndon words with 3 colors"); the third column is A072279 (``Dimension of n-th graded section of a certain Lie algebra"); and the fourth column is not yet an OEIS sequence.

\appendix

\section{Non-negativity of $\tau$}
\label{positivity_appendix}

In this Appendix, we prove the assertion from Section \ref{lattice_equivalent}: that if $\{\vec{m}_{1},\vec{m}_{2}\}$ and $\{\vec{m}_{1}',\vec{m}_{2}'\}$ are both {\it positive} integer bases for $\Lambda$ and, without loss of generality, we take the $\{\vec{m}_{1},\vec{m}_{2}\}$ parallelogram to be wider than the $\{\vec{m}_{1}',\vec{m}_{2}'\}$ parallelogram in the $\hat{e}_{\perp}$ direction, then the components $\{a,b,c,d\}$ of the $2\times2$ integer matrix $\tau$ are non-negative.

We can prove this as follows.  The positivity of the basis $\{\vec{m}_{1},\vec{m}_{2}\}$ implies that $m_{1}^{\parallel}$ and $m_{2}^{\parallel}$ are both positive, while $m_{1}^{\perp}$ and $m_{2}^{\perp}$ have opposite signs from one another; and, similarly, the positivity of the basis $\{\vec{m}_{1}',\vec{m}_{2}'\}$ implies that $m_{1}^{\parallel}{}'$ and $m_{2}^{\parallel}{}'$ are both positive, while $m_{1}^{\perp}{}'$ and $m_{2}^{\perp}{}'$ have opposite signs from one another.  Furthermore, for the purposes of this proof, we can restrict to the case $m_{1}^{\perp}<0$ and $m_{2}^{\perp}>0$ (since the other possibility corresponds to swapping $1\leftrightarrow 2$, which just swaps the columns of $\tau$, and does not affect the question of whether its components are all non-negative); and, similarly, we can restrict to the case $m_{1}^{\perp}{}'<0$ and $m_{2}^{\perp}{}'>0$ (since the other possibility corresponds to swapping $1'\leftrightarrow 2'$, which corresponds to swapping the rows of $\tau$, which again does not affect the question of whether its components are all non-negative).  With these restrictions, the requirement ${\rm det}(\tau)=\pm1$ reduces to the condition
\begin{subequations}
\begin{equation}
  \label{det_condition}
  {\rm det}(\tau)=1,
\end{equation}
and the requirement that the $\{\vec{m}_{1},\vec{m}_{2}\}$ parallogram is wider than the $\{\vec{m}_{1}',\vec{m}_{2}'\}$ parallelogram in the $\hat{e}_{\perp}$ direction reduces to the condition
\begin{equation}
  \label{width_condition}
  m_{2}^{\perp}-m_{1}^{\perp}>m_{2}^{\perp}{}'-m_{1}^{\perp}{}'.
\end{equation}
\end{subequations}

Now, using $\vec{m}_{1}'=a\vec{m}_{1}+b\vec{m}_{2}$, we see that the conditions $m_{1}^{\parallel}{}'>0$ and $m_{1}^{\perp}{}'<0$ become
\begin{subequations}
\begin{equation}
  a>-(m_{2}^{\parallel}/m_{1}^{\parallel})b\qquad{\rm and}\qquad
  a>-(m_{2}^{\perp}/m_{1}^{\perp})b.
\end{equation}
In other words, $a$ is greater than both $({\rm negative})\times b$ and $({\rm positive})\times b$, which is only possible if $a>0$.  Similarly, using $\vec{m}_{2}'=c\vec{m}_{1}+d\vec{m}_{2}$, the conditions $m_{2}^{\parallel}{}'>0$ and $m_{2}^{\perp}{}'>0$ become
\begin{equation}
   d>-(m_{1}^{\parallel}/m_{2}^{\parallel})c\qquad{\rm and}\qquad
   d>-(m_{1}^{\perp}/m_{2}^{\perp})c,
\end{equation}
\end{subequations}
which together imply $d>0$.

Next, conditions (\ref{det_condition}) and (\ref{width_condition}) may be rewritten, respectively, as
\begin{subequations}
  \begin{equation}
    \label{det_condition_new}
    bc=ad-1
  \end{equation}
  and 
  \begin{equation}
    \label{width_condition_new}
    b m_{2}^{\perp}-c m_{1}^{\perp}>(d-1)m_{2}^{\perp}-(a-1)m_{1}^{\perp}
  \end{equation}
\end{subequations}
Using the fact that $m_{2}^{\perp}$ is positive, $m_{1}^{\perp}$ is negative, while $a$ and $d$ are both positive integers, we see that Eqs.~(\ref{det_condition_new}) and (\ref{width_condition_new}) together imply that $b$ and $c$ are both non-negative.  

This completes the proof.

\section{Singular quasilattices}
\label{singular}

To describe quasilattices and their self-similarity transformations by a formula that continues to be precisely correct even in the singular case (see Sections \ref{
two_perspectives} and \ref{self_similar}), we must replace Eqs.~(\ref{xn_s_symmetric}, \ref{xn_s_asymmetric1}, \ref{xn_s_asymmetric2}) by
\begin{align}
    \frac{x_{n,s}}{\lambda_{\parallel}^{s}}
    &=\left(\left[\frac{n m_{2}^{\perp}-q_{0,s}^{\perp}}{m_{2}^{\perp}-m_{1}^{\perp}}\right]_{\sigma_{1,s}}\!\!\!\!+\frac{\sigma_{1,s}}{2}\right)m_{1}^{\parallel}\nonumber\\
    \label{xn_s_symmetric_prime}
    \tag{\ref{xn_s_symmetric}$'$} 
    &+\left(\left[\frac{n m_{1}^{\perp}-q_{0,s}^{\perp}}{m_{1}^{\perp}-m_{2}^{\perp}}\right]_{\sigma_{2,s}}\!\!\!\!+\frac{\sigma_{2,s}}{2}\right)m_{2}^{\parallel}-q_{0,s}^{\parallel} \\
    \label{xn_s_asymmetric1_prime}
    \tag{\ref{xn_s_asymmetric1}$'$}
    &=m_{1}^{\parallel}(n\!-\!\chi_{1,s}^{\parallel})\!+\!(m_{2}^{\parallel}\!-\!m_{1}^{\parallel})\!\left(\!\left[\kappa_{1}(n\!-\!\chi_{1,s}^{\perp})\right]_{\sigma_{2,s}}\!\!\!\!\!\!+\!
    \frac{\sigma_{2,s}}{2}\!\right) \\
    \label{xn_s_asymmetric2_prime}
    \tag{\ref{xn_s_asymmetric2}$'$}
    &=m_{2}^{\parallel}(n\!-\!\chi_{2,s}^{\parallel})\!+\!(m_{1}^{\parallel}\!-\!m_{2}^{\parallel})\!\left(\!\left[\kappa_{2}(n\!-\!\chi_{2,s}^{\perp})\right]_{\sigma_{1,s}}\!\!\!\!\!\!+\!
    \frac{\sigma_{1,s}}{2}\!\right)
\end{align}
where 
\begin{equation}
  \sigma_{1,s}\equiv({\rm det}\,\tau)^{s}\sigma_{1}\qquad{\rm and}\qquad
  \sigma_{2,s}\equiv({\rm det}\,\tau)^{s}\sigma_{2}.
\end{equation}
Here $\sigma_{1}$ and $\sigma_{2}$ are $\pm$ signs which may be regarded as independent in Eq.~(\ref{xn_s_symmetric_prime}), but are assumed to obey $\sigma_{1}=-\sigma_{2}$ in passing to Eqs.~(\ref{xn_s_asymmetric1_prime}, \ref{xn_s_asymmetric2_prime}).  Also note that we have introduced the notation
\begin{equation}
  [x]_{\sigma}=\left\{\begin{array}{cc} \lfloor x \rfloor & \qquad(\sigma=+) \\ \lceil x \rceil & \qquad(\sigma=-) \end{array}\right.
\end{equation}
where $\lfloor x\rfloor$ is the ``floor" of $x$ (the greatest integer $\leq x$) and $\lceil x \rceil$ is the ``roof" of $x$ (the least integer $\geq x$).  

In particular, note that the ``old" self-similarity transformation (\ref{xn_s_asymmetric1}, \ref{xn_s_asymmetric2}) corresponds to a fixed decoration/gluing rule {\it except} in the singular case (where the gluing/decoration rule hold {\it almost} everywhere, but is violated near the singular point in the quasilattice).  By contrast, the ``new" self-similarity transformation (\ref{xn_s_asymmetric1_prime}, \ref{xn_s_asymmetric2_prime}) corresponds to a fixed decoration/gluing rule that applies everywhere, even in the singular case.

\end{document}